\documentclass[pra,a4paper,twocolumn,floatfix]{revtex4}

\usepackage{graphicx}
\usepackage{amsmath, amsthm, amssymb}
\usepackage{subfigure}
\usepackage{comment}
\usepackage{url}
\usepackage{tikz}
\usepackage{youngtab}
\usepackage{young}

\newcommand{\C}{\mathbb{C}}

\newcommand{\ket}[1]{| #1 \rangle}
\newcommand{\bra}[1]{\langle #1|}

\newcommand{\tr}{\operatorname{tr}}

\newcommand{\be}{\begin{equation}}
\newcommand{\ee}{\end{equation}}
\newcommand{\bea}{\begin{eqnarray}}
\newcommand{\eea}{\end{eqnarray}}
\newcommand{\bes}{\begin{equation*}}
\newcommand{\ees}{\end{equation*}}
\newcommand{\beas}{\begin{eqnarray*}}
\newcommand{\eeas}{\end{eqnarray*}}

\newtheorem{thm}{Theorem}
\newtheorem*{thm*}{Theorem}
\newtheorem{cor}[thm]{Corollary}

\newtheorem{lem}[thm]{Lemma}
\newtheorem*{lem*}{Lemma}

\begin{document}

\date{\today}

\title{Symmetric functions of qubits in an unknown basis}

\author{Ashley Montanaro}
\affiliation{Department of Computer Science, University of Bristol, Woodland Road, Bristol, BS8 1UB, U.K.}
\email{montanar@cs.bris.ac.uk}

\begin{abstract}
Consider an $n$ qubit computational basis state corresponding to a bit string $x$, which has had an unknown local unitary applied to each qubit, and whose qubits have been reordered by an unknown permutation. We show that, given such a state with Hamming weight $|x| \le \lfloor n/2 \rfloor$, it is possible to reconstruct $|x|$ with success probability $1 - |x|/(n-|x|+1)$, and thus to compute any symmetric function of $x$. We give explicit algorithms for computing whether or not $|x|\ge t$ for some $t$, and for computing the parity of $x$, and show that these are essentially optimal. These results can be seen as generalisations of the swap test for comparing quantum states.
\end{abstract}

\maketitle


\section{Introduction}
\label{sec:introduction}

Consider the following scenario. Alice is a physicist who has just completed a long quantum computing experiment. Her quantum computer has produced an $n$ qubit state $\ket{x}$, corresponding to the bit string $x$. However, before she can measure the state to determine $x$, she is called away from the lab. In her absence, Eve sneaks in and sabotages the experiment. First, she applies an arbitrary local rotation to the qubits (the same rotation on each qubit); she then rearranges all the qubits in an arbitrary order.

It is clearly now hopeless for Alice to determine $x$ exactly. Indeed, she cannot even determine an individual bit of $x$ with any probability better than guessing. But what if she only needs to calculate $f(x)$, for some function $f$? Because of the arbitrary rearrangement of the qubits, she only has a chance of being able to compute symmetric functions, i.e.\ functions $f$ where $f(x)$ depends only on $|x|$, the Hamming weight of $x$. Also, because of the arbitrary local rotation, she can only compute functions $f$ where $f(x) = f(\bar{x})$, with $\bar{x}$ denoting bitwise negation. This is equivalent to imposing the constraint that $|x| \le \lfloor n/2 \rfloor$.

The purpose of this note is to show that Alice can in fact compute {\em any} $f$ that satisfies these constraints (with some probability, which may be low in the worst case). Indeed, we have the following result.

\begin{thm}
\label{thm:main}
Let $x$ be an $n$-bit string with $|x| \le \lfloor n/2 \rfloor$. Let $U$ be an unknown and arbitrary single qubit unitary operator, and $\sigma$ be an unknown and arbitrary permutation of $n$ qubits. Then there is a procedure which, given $\sigma(U^{\otimes n}\ket{x})$, outputs $|x|$ correctly with probability
\[ 1 - \frac{|x|}{n-|x|+1}. \]
Assuming that $|x|$ is distributed uniformly at random, for large $n$ this corresponds to an average probability of success of approximately $2(1-\ln 2) \approx 0.614$.
\end{thm}

Previous work has studied the closely related question of communicating without a shared reference frame \cite{bartlett03}. In this setting, Alice wishes to communicate some (classical or quantum) information to Bob by sending him $n$ qubits, but Bob does not know Alice's basis for each of the qubits. The results of \cite{bartlett03} show that, by encoding across multiple qubits, Alice can send Bob a number of bits that approaches $n$, in the large $n$ limit. By contrast, in the present work we do not allow prior encoding of Alice's information. Also note related previous work on the problem of computation in a hidden basis \cite{ioannou08}.

Theorem \ref{thm:main} can be used to obtain procedures for computing any symmetric function of $x$. These results can be seen as generalisations of the problem of determining whether $n$ qubits are all in the same state \cite{jex04,kada08}, which is in turn a generalisation of the question of determining equality of two qubits, which can be solved using the well-known {\em swap test} \cite{barenco97,buhrman01}. For example, we have the following results for the threshold function $\mbox{\sc Th}_t$ ($\mbox{\sc Th}_t(x) = 1 \Leftrightarrow |x| \ge t$) and the parity function ($\mbox{\sc Parity}(x) = \bigoplus_i x_i$).

\begin{cor}
\label{thm:threshold}
Let $x$ be an $n$-bit string with $|x| \le \lfloor n/2 \rfloor$. Let $U$ be an unknown and arbitrary single qubit unitary operator, and $\sigma$ be an unknown and arbitrary permutation of $n$ qubits. Then there is a procedure which, given $\sigma(U^{\otimes n}\ket{x})$, can compute $\mbox{\sc Th}_t(x)$ with success probability at least $1 - t/(n+1)$, and can compute $\mbox{\sc Parity}(x)$ with probability at least $1/2 + 1/(2(n+1))$.
\end{cor}

These success probabilities are essentially optimal, as we will show with the following theorem.

\begin{thm}
\label{thm:optimal}
Let $x$ be an $n$-bit string with $|x| \le \lfloor n/2 \rfloor$. Let $U$ be an unknown and arbitrary single qubit unitary operator, and $\sigma$ be an unknown and arbitrary permutation of $n$ qubits.  Let $f(x)$ be some function such that $f(k+1) \neq f(k)$ for some $0 \le k < \lfloor n/2 \rfloor$. Then any procedure that computes $f(x)$ given access to $\sigma(U^{\otimes n}\ket{x})$ succeeds with probability at most $1-(k+1)/(2(n-k))$ in the worst case.
\end{thm}

This implies that, for example, one can determine whether or not $|x|=0$ very effectively, but distinguishing $|x|=n/2$ from $|x|=n/2-1$ is hard. Interestingly, this phenomenon also occurs in the study of quantum and classical query complexity \cite{beals01}, and more generally in approximation theory \cite[Section 3.4]{petrushev87}.


\section{The symmetric group and weak Schur sampling}

The results in this note will be proven using some basic representation theory of the symmetric group $S_n$, which we now outline. Conjugacy classes of $S_n$ are labelled by partitions $\lambda \vdash n$. Each partition $\lambda$ containing $k$ parts can be written as a non-increasing sequence of positive integers $(\lambda_1,\dots,\lambda_k)$, and expressed as a {\em Young diagram}. The diagram corresponding to the partition $\lambda$ is a collection of boxes arranged in left-justified rows, where the number of boxes in row $i$ is given by $\lambda_i$. Irreducible representations (irreps) of the symmetric group are thus in one-to-one correspondence with Young diagrams. We let $\lambda$ denote both a partition and its corresponding diagram, and $V_\lambda$ denote the corresponding irrep.

\begin{figure}[ht]
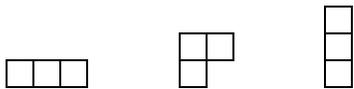

\begin{center}
\yng(3) \hspace{1cm} \yng(2,1) \hspace{1cm} \yng(1,1,1)
\end{center}
\caption{The irreducible representations (3), (2,1), (1,1,1) of the group $S_3$.}
\end{figure}

We think of the input to our problem as an $n$-qubit state $\ket{x}$ in a known basis, to which an unknown, and arbitrary, tensor product unitary $U^{\otimes n}$ has been applied, followed by an unknown permutation of the qubits $\sigma$. In order to take advantage of these symmetries, we will use {\em Schur-Weyl duality}. This states that the space of $n$ qudits decomposes into a direct sum of tensor products of subspaces corresponding to irreps of the symmetric and unitary groups, as follows:
\be
\label{eqn:schur}
(\C^d)^{\otimes n} \cong \bigoplus_{\lambda \vdash n} \mathcal{P}_\lambda \otimes \mathcal{Q}_\lambda^d,
\ee
where $\mathcal{P}_\lambda$ and $\mathcal{Q}_\lambda^d$ correspond to irreps of $S_n$ and $U_d$, respectively. For good introductions to Schur-Weyl duality in the context of quantum information theory, see the theses \cite{harrow05} and \cite{christandl06}.

The Schur transform \cite{harrow05,bacon06,bacon07} performs an implementation of this decomposition, mapping a state in the computational basis to one of the form $\ket{\lambda}\ket{p}\ket{q}$. In this case, as we are indifferent to permutations of the subsystems and local unitaries on each subsystem, we will only measure the $\ket{\lambda}$ register. This is known as {\em weak Schur sampling} \cite{childs07c}. The projector onto a given value of $\lambda$ is given by (see \cite{childs07c} or \cite[Theorem 8]{serre77})
\be \label{eqn:plambda} P_\lambda = \frac{d_\lambda}{n!} \sum_{\pi \in S_n} \chi_\lambda(\pi) D(\pi), \ee
where $d_\lambda$ is the dimension of the irrep $V_\lambda$, $\chi_\lambda$ is the character $\tr V_\lambda$, and $D$ is the defining representation of $S_n$ that acts by permuting the $n$ subsystems,
\[ D(\pi) \ket{i_1} \cdots \ket{i_n} = \ket{i_{\pi^{-1}(1)}} \cdots \ket{i_{\pi^{-1}(n)}}. \]
It is unnecessary to perform the full Schur transform to measure $\lambda$; it suffices to use the quantum Fourier transform over the symmetric group $S_n$, in a procedure known as generalised phase estimation \cite{harrow05,childs07c}, which can be seen as a generalisation of the swap test \cite{barenco97,buhrman01}. To perform generalised phase estimation on an $n$ qubit state $\rho$, one first prepares an ancilla register in the state $\frac{1}{n!} \sum_{\pi \in S_n} \ket{\pi}$. This register is used to control a conditional permutation of the subsystems of $\rho$, which is followed by an inverse quantum Fourier transform (over $S_n$) on the ancilla register. Measuring the ancilla gives a value of $\lambda$ with probability $\tr(P_\lambda \rho)$. This whole procedure can be performed efficiently, i.e.\ in time polynomial in $n$ \cite{beals97}.

Let $\ket{x}$ be the input state and assume that $x$ has Hamming weight $k \le \lfloor n/2 \rfloor$. Letting $\sigma$ be an arbitrary permutation of $n$ qubits and $U$ be an arbitrary local unitary, we now compute $\tr(P_\lambda\,D(\sigma) U^{\otimes n} \ket{x} \bra{x} (U^\dag)^{\otimes n}D(\sigma)^{\dag})$. As $P_\lambda$ commutes with local unitaries and permutations \cite{childs07c}, this is equal to $\tr(P_\lambda \ket{x} \bra{x})$. Invariance under permutation also implies that the probability of obtaining a given outcome $\lambda$ depends only on the Hamming weight of $x$. The following crucial lemma allows us to write down exactly what these probabilities are.

\begin{lem}
\label{lem:probs}
Let $\Pr[\ell|k]$ denote the probability of getting the measurement outcome corresponding to the partition $(n-\ell,\ell)$ when performing weak Schur sampling on an $n$-bit string with Hamming weight $k$. Then, if $\ell > k$, $\Pr[\ell|k] = 0$. Otherwise,
\[ \Pr[\ell|k] = \frac{\binom{n}{\ell}-\binom{n}{\ell-1}}{\binom{n}{k}}. \]
In particular, $\sum_{\ell=0}^k \Pr[\ell|k] = 1$.
\end{lem}

Before we prove this lemma, we show that it implies the results stated in Section \ref{sec:introduction}. In the case of Theorem \ref{thm:main}, we give an explicit algorithm that achieves the success probability required by the theorem:

\begin{enumerate}
\item Perform weak Schur sampling, obtaining outcome $\lambda = (n-\ell,\ell)$.
\item Output the guess that $k = \ell$.
\end{enumerate}

It is clear that this procedure, which we will term the {\em standard algorithm}, will output the correct answer with probability $\Pr[k|k] = 1 - k/(n-k+1)$.

One might consider more complicated strategies for inferring $k$ from weak Schur sampling. A general inference strategy can be expressed as a matrix $O$, where $O_{k\ell} = \Pr[\mbox{output $k$}|\mbox{get outcome $(n-\ell,\ell)$}]$, and $\sum_k O_{k\ell} = 1$ for all $\ell$. If one wishes to maximise the worst-case probability of outputting the correct value of $k$, for example, it is required to find an $O$ that maximises
\[ \min_k \left( \sum_{\ell=0}^k O_{k \ell} \left( \frac{\binom{n}{\ell}-\binom{n}{\ell-1}}{\binom{n}{k}} \right) \right). \]
This is a linear programming problem and can be solved exactly for small $n$, although we do not know a closed form for the solution for general $n$.

An alternative setting for the inference problem is the Bayesian scenario where one maximises the probability of success assuming an a priori probability distribution on $k$ (see \cite{mackay03} for a comprehensive introduction to Bayesian inference). Letting $\{p_k\}$ denote this probability distribution, the problem is to maximise
\beas
P_{succ} &=& \sum_{k=0}^{\lfloor n/2 \rfloor} p_k \left( \sum_{\ell=0}^k O_{k \ell} \left( \frac{\binom{n}{\ell}-\binom{n}{\ell-1}}{\binom{n}{k}} \right) \right)\\
&=& \sum_{\ell=0}^{\lfloor n/2 \rfloor} \left( \binom{n}{\ell}-\binom{n}{\ell-1} \right) \left( \sum_{k=\ell}^{\lfloor n/2 \rfloor} O_{k \ell} \frac{p_k}{\binom{n}{k}}  \right).
\eeas
This is clearly maximised by taking $O_{k\ell} = 1$ for $k = \max_{k'} p_{k'}/\binom{n}{k'}$, and $O_{k\ell} = 0$ otherwise. In the particularly natural case where we assume that the a priori distribution on $k$ is uniform, this maximisation in fact shows that the standard algorithm is optimal, and gives an average probability of success of
\[ \frac{1}{\lfloor n/2 \rfloor + 1} \sum_{k=0}^{\lfloor n/2 \rfloor} 1 - k/(n-k+1). \]
For large $n$, this can be estimated as
\begin{gather*}
\hspace{-2cm}
1 - \frac{2}{n} \int_0^{n/2} k/(n-k+1)\,dk\\
\begin{split}
&= 2\left(1-\frac{(n+1)}{n} \ln \left( \frac{n+1}{n/2+1}\right) \right)\\
&\approx 2(1-\ln 2) \approx 0.614.
\end{split}
\end{gather*}
What about the scenario of Corollary \ref{thm:threshold}, where we only want to compute some function $f(k)$, rather than to output $k$? It is natural to try to produce an algorithm with high success probability for all $0 \le k \le \lfloor n/2 \rfloor$. Again, if one attempts to maximise this worst-case probability over all strategies that consist of performing weak Schur sampling and attempting to infer $f(k)$ from the result, one is led to a linear programming problem for which we do not know a closed form solution. A more straightforward approach is to guess $k$ using the standard algorithm (call this guess $\widetilde{k}$), and then to output $f(\widetilde{k})$.

The probability that this gives the right answer can easily be calculated for threshold functions $\mbox{\sc Th}_t$. Assuming $k \ge t$,
\beas
\Pr[f(\widetilde{k})\neq f(k)] &=& \sum_{\ell,f(\ell)\neq f(k)} \Pr[\ell|k]\\
&=& \frac{1}{\binom{n}{k}} \sum_{\ell=0}^{t-1} \left(\binom{n}{\ell} - \binom{n}{\ell-1}\right)=\frac{\binom{n}{t-1}}{\binom{n}{k}}.
\eeas
This is clearly maximised by $k=t$, giving a failure probability of at most $t/(n-t+1)$. On the other hand, if $k<t$, note that this algorithm succeeds with certainty, as $\Pr[\ell|k]=0$ for $\ell > k$. We thus have a probabilistic algorithm that computes the threshold function $\mbox{\sc Th}_t$ with one-sided error. This can be modified to give an algorithm with small worst-case probability of error, as follows.

Consider any procedure that attempts to compute an arbitrary boolean function $f(k)$ from $f(\widetilde{k})$, where $k$ is picked to minimise the probability that $f(k) = f(\widetilde{k})$. Such a procedure can be parametrised by two probabilities $q_0$, $q_1$, where $q_0$ is the probability that the procedure outputs 0, given that $f(\widetilde{k}) = 0$, and $q_1$ is the probability of outputting 0, given that $f(\widetilde{k}) = 1$. Let $p_i = \Pr [f(\widetilde{k})=0|f(k)=i]$ for $i \in \{0,1\}$, and assume that $p_0 \ge p_1$. Then the probability of success of such a procedure (in the worst case) is at least
\[ \min \{ q_0\,p_0 + q_1 (1-p_0), (1-q_0)p_1 + (1-q_1)(1-p_1) \}. \]
We pick $q_1$ such that these two values are equal, which gives
\[ q_1 = \frac{1-q_0(p_0+p_1)}{2-p_0-p_1}. \]
Our goal is to maximise the corresponding expression for the probability of success,
\[ q_0\,p_0 + q_1 (1-p_0) = \frac{1+q_0(p_0-p_1)-p_0}{2-p_0-p_1}, \]
over $q_0$, while still obeying the constraints $0 \le q_0, q_1 \le 1$. This is straightforward and gives the answer
\[ q_0 = \min \{ \frac{1}{p_0+p_1}, 1 \}, q_1 = \max \{ 0, \frac{1-p_0-p_1}{2-p_0-p_1} \}, \]
which corresponds to a maximum worst-case probability of success of $p_0/(p_0+p_1)$ when $p_0 + p_1 \ge 1$, and $(1-p_1)/(2-p_0-p_1)$ when $p_0 + p_1 \le 1$.

Applying this result to the threshold function $\mbox{\sc Th}_t$, where $p_0 = 1$ and $p_1 = t/(n-t+1)$, it can easily be seen that we obtain a two-sided error algorithm that always succeeds with probability at least $1 - t/(n+1)$, as stated in Corollary \ref{thm:threshold}.

In the case of the {\sc Parity} function, we can calculate the probability that $f(\widetilde{k}) \neq f(k)$ from
\begin{gather*}
\hspace{-4cm}\Pr[\widetilde{k}\mbox{ is even}]-\Pr[\widetilde{k}\mbox{ is odd}]\\
\begin{split}
&= \frac{1}{\binom{n}{k}} \sum_{\ell=0}^k (-1)^{\ell} \left( \binom{n}{\ell} - \binom{n}{\ell-1} \right)\\
&= (-1)^k \left(1 - \frac{2k}{n}\right).
\end{split}
\end{gather*}
It is then immediate that
\begin{equation*}
\Pr[\widetilde{k}\mbox{ is even}] = \left\{ 
\begin{aligned}
 1-k/n &\,\,\text{($k$ even)} \\
 k/n &\,\,\text{($k$ odd),}
  \end{aligned}
\right.
\end{equation*}
and of course $\Pr[\widetilde{k}\mbox{ is odd}] = 1-\Pr[\widetilde{k}\mbox{ is even}]$. Assume that $n$ is even and $n/2$ is also even (the cases where $n$ or $n/2$ is odd are analogous). This implies that, for $k$ even, we have $\Pr[f(\widetilde{k})=f(k)] \ge 1/2$, and for $k$ odd, $\Pr[f(\widetilde{k})=f(k)] \ge 1/2 + 1/n$. We can use the same technique as before to get an algorithm that succeeds with probability at least $1/2 + 1/(2(n+1))$ in the worst case.


\section{Limits on success probability}

To prove Theorem \ref{thm:optimal}, and hence to show that these algorithms are almost optimal, consider the restricted problem of distinguishing a bit-string $x$ with weight $k$ from one with weight $k+1$, as is required to compute a symmetric function $f$ where $f(k) \neq f(k+1)$.

We first argue that any procedure for computing $f$ might as well simply consist of weak Schur sampling and post-processing the results. Imagine that an adversary, as in the scenario of Section \ref{sec:introduction}, has performed a random permutation and a random local rotation of each qubit on the initial state $\ket{x}$. Then, from Alice's perspective, the resulting state looks like
\[ \rho = \frac{1}{n!} \int_{U} dU\,U^{\otimes n} \sum_{\sigma \in S_n} D(\sigma) \ket{x}\bra{x} D(\sigma)^{\dag} (U^{\dag})^{\otimes n}, \]
which is equal to $\sum_{\lambda \vdash n} k_\lambda P_\lambda$ for some coefficients $\{k_\lambda\}$. (This follows from the decomposition of eqn.\ (\ref{eqn:schur}) and Schur's Lemma, using the fact that $\rho$ commutes with all permutations and local unitaries.) This implies that, without loss of generality, a measurement strategy can be taken as consisting of measuring $\lambda$ and performing some classical post-processing.

Let $p_k(\ell)$ denote the probability distribution over partitions $(n-\ell,\ell)$ obtained by performing weak Schur sampling on an input with weight $k$. We calculate the $\ell_1$ distance between the distributions $p_k$, $p_{k+1}$ for arbitrary $0 \le k < \lfloor n/2 \rfloor$:
\beas
\|p_k - p_{k+1}\|_1 &=& \frac{\binom{n}{k+1}-\binom{n}{k}}{\binom{n}{k+1}}\\
&& \hspace{-2cm} + \sum_{\ell=0}^{k} \left(\binom{n}{\ell} - \binom{n}{\ell-1}\right) \left(\frac{1}{\binom{n}{k}} - \frac{1}{\binom{n}{k+1}}\right)\\
&=& 2\left(1-\frac{\binom{n}{k}}{\binom{n}{k+1}} \right) = 2\left(\frac{n-2k-1}{n-k}\right).
\eeas
Using standard results on distinguishing probability distributions, this distance puts an upper bound on the probability of success of any algorithm attempting to distinguish between weights $k$ and $k+1$, and implies that the above algorithms are asymptotically optimal. We finally turn to the proof of Lemma \ref{lem:probs}.


\section{Proof of Lemma \ref{lem:probs}}

Let $\lambda$ be the partition $(n-\ell,\ell)$ and let $x$ be a bit-string with Hamming weight $k \le \lfloor n/2 \rfloor$, assuming without loss of generality that $\ket{x} = \ket{1\cdots 1 0 \cdots 0}$, where the first $k$ bits of $x$ are 1 and the last $n-k$ are 0. We now calculate $\Pr[\ell|k] = \tr(P_\lambda \ket{x} \bra{x})$ using (\ref{eqn:plambda}). It is easy to see that $\tr(D(\pi) \ket{x} \bra{x})=0$ unless $\pi$ leaves the bit-string $x$ unchanged, in which case $\tr(D(\pi) \ket{x} \bra{x})=1$. All such permutations $\pi$ can be decomposed as a direct product of a permutation of the first $k$ bits, and a permutation of the last $n-k$ bits. This implies that
\[ \tr(P_\lambda \ket{x} \bra{x}) = \frac{d_\lambda}{n!} \tr\left(\sum_{\pi \in S_k \times S_{n-k}} V_\lambda(\pi)\right). \]
We first calculate the sum over the group $S_k \times S_{n-k}$. Note that the representation $V_\lambda$, while irreducible over $S_n$, is not necessarily irreducible over $S_k \times S_{n-k}$, but may split into a direct sum of irreps. The following simple lemma, which can be proven using Schur's Lemma, shows that only the trivial irrep is of interest.
\begin{lem}
Let $V_\lambda$ be an irreducible representation of a finite group $G$. Then
\begin{equation*}
\sum_{g \in G} V_\lambda(g) = \left\{ 
\begin{aligned}
 |G| &\,\,\text{ if $V_\lambda$ is the trivial irrep} \\
 0 &\,\,\text{ otherwise.}
  \end{aligned}
\right.
\end{equation*}
\end{lem}
Each occurence of the trivial irrep in the decomposition of the representation $V_\lambda$ over $S_k \times S_{n-k}$ will thus give a contribution of $k!(n-k)!$ to the sum; all other irreps will contribute nothing. This number of occurrences can be calculated using a special case of the {\em Littlewood-Richardson rule} known as Pieri's formula \cite{fulton91}.

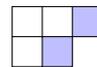
\begin{figure}[ht]
\begin{center}
\begin{tikzpicture}[scale=0.4]
\fill[blue!25] (2,1) rectangle (3,2);
\fill[blue!25] (1,0) rectangle (2,1);
\draw (1,2) -- (1,0) -- (0,0) -- (0,2) -- (3,2) -- (3,1) -- (0,1);
\draw (1,0) -- (2,0) -- (2,2);
\end{tikzpicture}
\end{center}
\caption{Expanding the diagram $(2,1)$ to $(3,2)$ using the diagram $(2)$.}
\label{fig:expand}
\end{figure}

Let $\mu$ be the diagram corresponding to the trivial irrep of $S_k$, for some $k$. Then, for any $\lambda$ and $\nu$, we define the Littlewood-Richardson number $N_{\lambda \mu \nu}$ as the number of ways that $\lambda$ can be expanded to $\nu$ by adding $k$ boxes to $\lambda$, under the constraint that at most one new box is added to each column. See Figure \ref{fig:expand} for an illustration of this process, and note that $N_{\lambda \mu \nu}$ is always either 0 or 1 (though this does not remain true in the more general setting where $\mu$ can be arbitrary; then the rule is more complicated).

Littlewood-Richardson numbers are relevant because of the following theorem \cite{fulton91}.

\begin{thm}
Let $\mu$ be the partition $(n-k)$. The multiplicity of the irrep $V_\lambda \otimes V_\mu$ in the restriction of the irrep $V_\nu$ from $S_n$ to $S_k \times S_{n-k}$ is equal to $N_{\lambda \mu \nu}$.
\end{thm}

As we are only interested in expansions of the trivial irrep $(k)$ by another trivial irrep $(n-k)$, this multiplicity is particularly simple to calculate. Let $\nu$ be a partition of $n$, $\lambda$ be the partition $(k)$, and $\mu$ be the partition $(n-k)$. Then, if $\nu$ has more than two parts, $N_{\lambda \mu \nu} = 0$. (This was expected anyway because each of the $n$ subsystems we are dealing with has dimension 2.) If $\nu$ has two parts, express it as $(n-\ell, \ell)$. Then, if $k < \ell$, $N_{\lambda \mu \nu} = 0$. Otherwise, $N_{\lambda \mu \nu} = 1$.

\begin{figure}[ht]
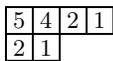

\begin{center}
\young(5421,21)
\end{center}
\caption{Hook-lengths of the cells in the diagram (4,2).}
\label{fig:hooklengths}
\end{figure}

This deals with the sum; to finish the calculation, we need to find the dimension $d_\lambda$. This can be evaluated using the famous {\em hook-length formula} \cite{fulton91}. Let $x$ be a box in a Young diagram. Then the hook-length $h(x)$ is defined as the total number of boxes in the same row and to the right of $x$, plus the total number in the same column and below $x$, plus 1 (for $x$ itself). See Figure \ref{fig:hooklengths} for an illustration.

The hook-length formula states that
\[ d_\lambda = \frac{n!}{\prod_{x \in \lambda} h(x)}. \]
As we only need calculate $d_\lambda$ for partitions $\lambda = (n-\ell,\ell)$, this formula is particularly simple, and gives
\[ d_\lambda = \frac{n!(n-2\ell+1)}{\ell!(n-\ell+1)!} = \binom{n}{\ell} \left( \frac{n-2\ell+1}{n-\ell+1} \right). \]
To sum up, we have, for $\ell \le k$,

\beas
\Pr[\ell|k] &=& \frac{d_\lambda}{n!} \tr\left(\sum_{\pi \in S_k \times S_{n-k}} V_\lambda(\pi)\right)\\
&=& \frac{1}{n!} \binom{n}{\ell} \left( \frac{n-2\ell+1}{n-\ell+1} \right) k!(n-k)!\\
&=& \frac{\binom{n}{\ell}}{\binom{n}{k}} \left( \frac{n-2\ell+1}{n-\ell+1} \right) = \frac{\binom{n}{\ell}-\binom{n}{\ell-1}}{\binom{n}{k}},
\eeas
where the last step is a binomial coefficient identity that can be verified directly. For $\ell > k$, by Pieri's formula the sum over $S_k \times S_{n-k}$ is zero. This implies that $\Pr[\ell|k]=0$ in this case, and completes the proof.


\section{Conclusion}

We conclude that it is possible to compute symmetric functions of an $n$ qubit state $\ket{x}$, even if a malicious adversary has applied an arbitrary local rotation and an arbitrary permutation to the state, even without any prior encoding of $x$. This is in (perhaps surprising) contrast to the fact that any individual bit of $x$ cannot be retrieved.

\vfill
\section*{Acknowledgements}

This work was supported by the EC-FP6-STREP network QICS, and was partly carried out during a visit to the Perimeter Institute for Theoretical Physics. I would like to thank Aram Harrow, Richard Low and Tobias Osborne for helpful discussions on the subject of this work, and an anonymous referee for helpful comments which improved the paper.



\end{document}